\documentclass[sigconf]{acmart}

\setcopyright{none}
\renewcommand\footnotetextcopyrightpermission[1]{}
\usepackage{enumitem,kantlipsum}

\AtBeginDocument{%
  \providecommand\BibTeX{{%
    \normalfont B\kern-0.5em{\scshape i\kern-0.25em b}\kern-0.8em\TeX}}}

\acmConference[arxiv]{Make sure to enter the correct
  conference title from your rights confirmation emai}{November,
  2022}{}
\acmPrice{15.00}
\acmISBN{978-1-4503-XXXX-X/18/06}

\begin{document}

\title{BCIM: Efficient Implementation of Binary Neural Network Based on Computation in Memory }

\author{Mahdi Zahedi}
\email{m.z.zahedi@tudelft.nl}
\orcid{0000-0002-7602-5066}
\affiliation{%
  \institution{Delft University of Technology}
  \streetaddress{Mekelweg 5}
  \city{Delft}
  \state{South Holland}
  \country{The Netherlands}
  \postcode{2628 CD}
}

\author{Taha Shahroodi}
\email{T.Shahroodi@tudelft.nl}
\affiliation{%
  \institution{Delft University of Technology}
  \streetaddress{Mekelweg 5}
  \city{Delft}
  \state{South Holland}
  \country{The Netherlands}
  \postcode{2628 CD}
}

\author{Stephan Wong}
\email{j.s.s.m.wong@tudelft.nl}
\affiliation{%
  \institution{Delft University of Technology}
  \streetaddress{Mekelweg 5}
  \city{Delft}
  \state{South Holland}
  \country{The Netherlands}
  \postcode{2628 CD}
}

\author{Said Hamdioui}
\email{s.hamdioui@tudelft.nl}
\affiliation{%
  \institution{Delft University of Technology}
  \streetaddress{Mekelweg 5}
  \city{Delft}
  \state{South Holland}
  \country{The Netherlands}
  \postcode{2628 CD}
}


\begin{abstract}
  Applications of Binary Neural Networks (BNNs) are promising for embedded systems with hard constraints on computing power. Contrary to conventional neural networks with the floating-point datatype, BNNs use binarized weights and activations which additionally reduces memory requirements. Memristors, emerging non-volatile memory devices, show great potential as the target implementation platform for BNNs by integrating storage and compute units. The energy and performance improvements are mainly due to 1) accelerating matrix-matrix multiplication as the main kernel for BNNs, 2) diminishing memory bottleneck in von-Neumann architectures, 3) and bringing massive parallelization. However, the efficiency of this hardware highly depends on how the network is mapped and executed on these devices. In this paper, we propose an efficient implementation of XNOR-based BNN to maximize parallelization while using a simple sensing scheme to generate activation values. Besides, a new mapping is introduced to minimize the overhead of data communication between convolution layers mapped to different memristor crossbars. This comes with extensive analytical and simulation-based analysis to evaluate the implication of different design choices considering the accuracy of the network. The results show that our approach achieves up to $10\times$ energy-saving and $100\times$ improvement in latency compared to the state-of-the-art in-memory hardware design.               
\end{abstract}





\maketitle

\section{Introduction}
\label{section:Introduction}


Machine learning algorithms and specifically Deep Neural Networks (DNNs) have pushed the state-of-the-art designs and become prominent in a variety of applications, including, but not limited to language processing \cite{DNN_language}, object recognition \cite{DNN_object}, and image classification \cite{DNN_image1,DNN_image2}. Designing larger networks and the ability to train them with advanced algorithms was the main driver to enable performing complex applications for several years. Besides advanced algorithms, hardware implementation and its challenges play a major role in the deployment and development of DNN applications specifically for embedded systems. One of the main hardware challenges of NN is the large data set (weights) that has to be stored and performed computation on (memory wall) \cite{BNN_survey}. Neural networks usually use floating-point computation which requires large storage and many resources. As a response to this challenge, Binary Neural Networks (BNNs), where the weights and activation values are binarized, receives more attention from researchers \cite{review_BNN}. A BNN reduces memory consumption and simplified computations which leads to a higher energy-efficient system. However, this efficiency highly depends on the implementation of the network considering the hard constraints of embedded systems. Therefore, considerable research is required to ensure the effectiveness of BNNs for state-of-the-art applications.     

In recent years, many works have been published with a focus on the algorithmic optimizations of BNNs. Minimizing quantization error \cite{xnor_net,bnn_quantize,bnn_quantize2}, improving the network loss function \cite{bnn_loss1,bnn_loss2}, and reducing gradient error \cite{gradient1,gradient2+} have been the main topics of interest. From the hardware perspective, besides using traditional systems (CPU, GPU, and FPGA) \cite{BNN_impl1,BNN_impl2,BNN_impl3}, memristor-based BNN accelerators are getting more attention due to reduced communication overhead, as the main bottleneck in traditional computing systems, by deploying computation in memory as a concept \cite{Said_app}. However, using memristors for signed numbers is challenging. From this perspective, existing works can be classified into hardware or algorithmic solutions. Positive and negative values can be mapped to different memristors \cite{highly_parallel_NN,BCNN_RRAM,configurable_neurons}. Other approaches are considering one- \cite{BNN_ref_column1} or two-column reference memristors \cite{highly_robust} while the weights and activation are presented as unsigned numbers. In general, these approaches require more devices, increase design complexity, and reduce energy/performance-efficiency of the system.  As an algorithmic solution, the signed multiply-and-accumulate which is the main operation in BNNs can be converted to XNOR operations \cite{xnor_net}. Using this method, it was proposed to use memristors as an activation function \cite{BNN_SA_is_memristor}, but this induces endurance, energy, and performance issues due to the excessive programming. To ensure the accuracy of XNOR operations against device variation, a new memristor crossbar structure based on differential sensing is used \cite{BNN_differential_SA}. However, XNOR operations are forced to be performed sequentially due to the sensing mechanism. All these overheads drive researchers to explore efficient implementations of BNN specifically for embedded systems. This is inevitable considering advanced workloads demanding more energy and computing times.

This work advances the state-of-the-art by proposing an efficient implementation of BNNs. In this design, we mimic the functionality of ADC and the required following digital processing by a Sense Amplifier (SA) while it allows simultaneous row activation to maximize resource utilization on the crossbar and enhance the performance. Extensive analytical analysis and simulations are performed to ensure the accuracy of the design considering the scenarios where the design behaves as an approximation. The effect of the number of references for the SA and the distance between the values of the references are studied. Furthermore, we minimize data communication between layers by proposing a novel mapping of the weights and activation values into the crossbar and its input buffer, respectively. 
Finally, we investigate the efficiency of our approach on different network structures in terms of accuracy, energy, and performance by developing our PyTorch-based simulation platform intended to be open-sourced. The platform can mimic the behavior of the crossbar and allows for more characteristics and non-idealities to be integrated and explored for different networks.              
In summary, this paper presents the following main contributions:  \vspace{-0.1cm}
\begin{itemize}
    \item We propose an energy-efficient and highly parallel implementation of XNOR-based BNNs where the functionality of ADC and the required post-processing after that is modeled by a SA.      
    \item We perform extensive analytical as well as simulation-based analysis where the proposed implementation behaves as an approximation in order to comprehend the implication of SA on the accuracy of the design.       
    \item We present an efficient mapping of the weights and activation values to improve data utilization and minimize the number of communication between network layers. The technique is general and can be applied to non-binary networks as well.  
\end{itemize}

\section{Preliminary}
\label{section:Preliminary}
\begin{figure}[t]
\centering
  \includegraphics[width=0.8\linewidth]{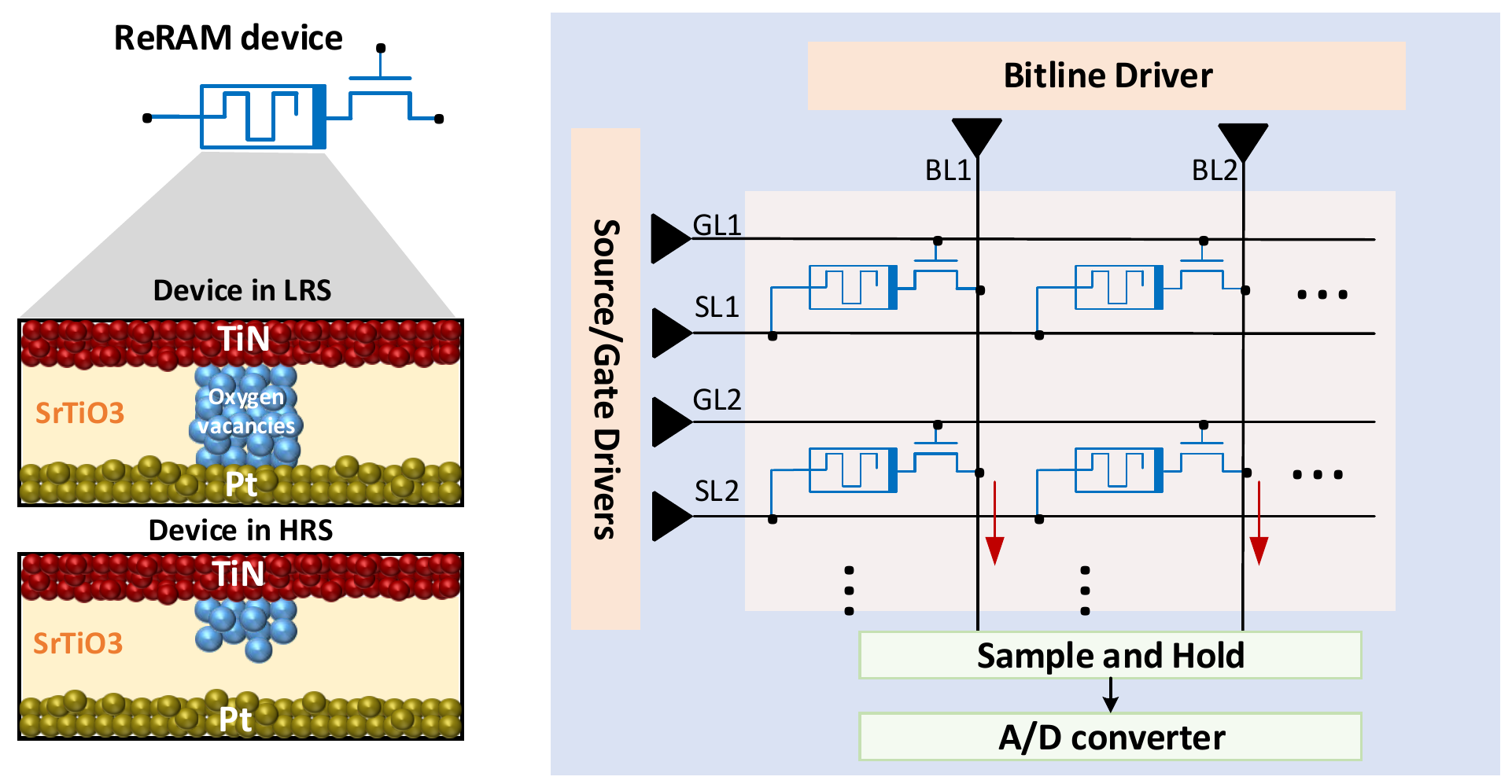}
  \caption{Memristor ReRAM device behavior in LRS and HRS mode as well as a crossbar structure}
  \label{fig:memristor}
\end{figure}
In this section, two topics are covered. First, we provide background information about memristor devices and the operations supported in a crossbar array. Second, we explain briefly the basics of binary neural networks.  

\subsection{Memristor devices}
Contrary to charge-based memories, memristor devices are categorized as non-volatile memory where data can be represented as a low resistive state (LRS) and a high resistive state (HRS) by application of appropriate voltage signals. Many technologies can be used to build these devices such as Resistive Random-Access Memory (ReRAM or RRAM) or Phase-Change Memory (PCM) \cite{RRAM_intel,Abu_review}. As an example, ReRAM consists of a metal-insulator-metal stack where the device is set and reset by changing the polarity of the programming voltage to form or dissolve the conducting filament (Figure \ref{fig:memristor}). The resistance level indicates the logic value intended to be stored in the device. In order to read the device without disturbance, a small voltage should be applied and the current (voltage) through (across) the device should be sensed. Figure \ref{fig:memristor} depicts a 1T1R crossbar structure where three drivers are employed to program or read the devices. In the case of read or computational operations, the current passes through the bitline is sensed, and converted to digital domain using a SA or ADC. The main computational operations that can be performed on the crossbar include addition, logical operations, and Matrix-Matrix Multiplication (MMM). Besides the capabilities of co-locating computation and storage together, huge parallelism can be achieved within a single memory array (crossbar and its periphery) as well as at the inter-array level. These are the main drivers that attracts researchers to exploit this concept for different state-of-the-art applications \cite{Said_app}.    

\subsection{Fundamentals of Binary Neural Networks}
Nowadays, deeper neural networks have been developed to be able to perform advanced tasks such as complex classification or image segmentation. However, implementing these networks in embedded platforms with limited storage and computation units is challenging specifically in consideration of strict energy/performance constraints. Despite conventional neural networks with high precision datatypes, in BNNs, weights and activations are binarized to make the network extremely compact. Equation \ref{eq:sign} shows a simple binarization rule that can be applied to both activations (input tensors) and weights where $B_{\omega}$ and $B_{\textit{I}}$ are the binarized weights and input tensors, respectively. The binarization not only saves on the storage usage, but also reduces the expensive multiply-accumulate operation to a simple addition.         

\begin{equation}
  \label{eq:sign}
  \begin{split}
  B_{\omega} =
    \begin{cases}
      +1 & if \; \omega \geq 0\\
      -1 & if \; \omega < 0
    \end{cases}  
    B_{\textit{I}} =
    \begin{cases}
      +1 & if \; \textit{I} \geq 0\\
      -1 & if \; \textit{I} < 0
    \end{cases}
  \end{split}
\end{equation}

\begin{figure}[t]
\centering
  \includegraphics[width=0.7\linewidth]{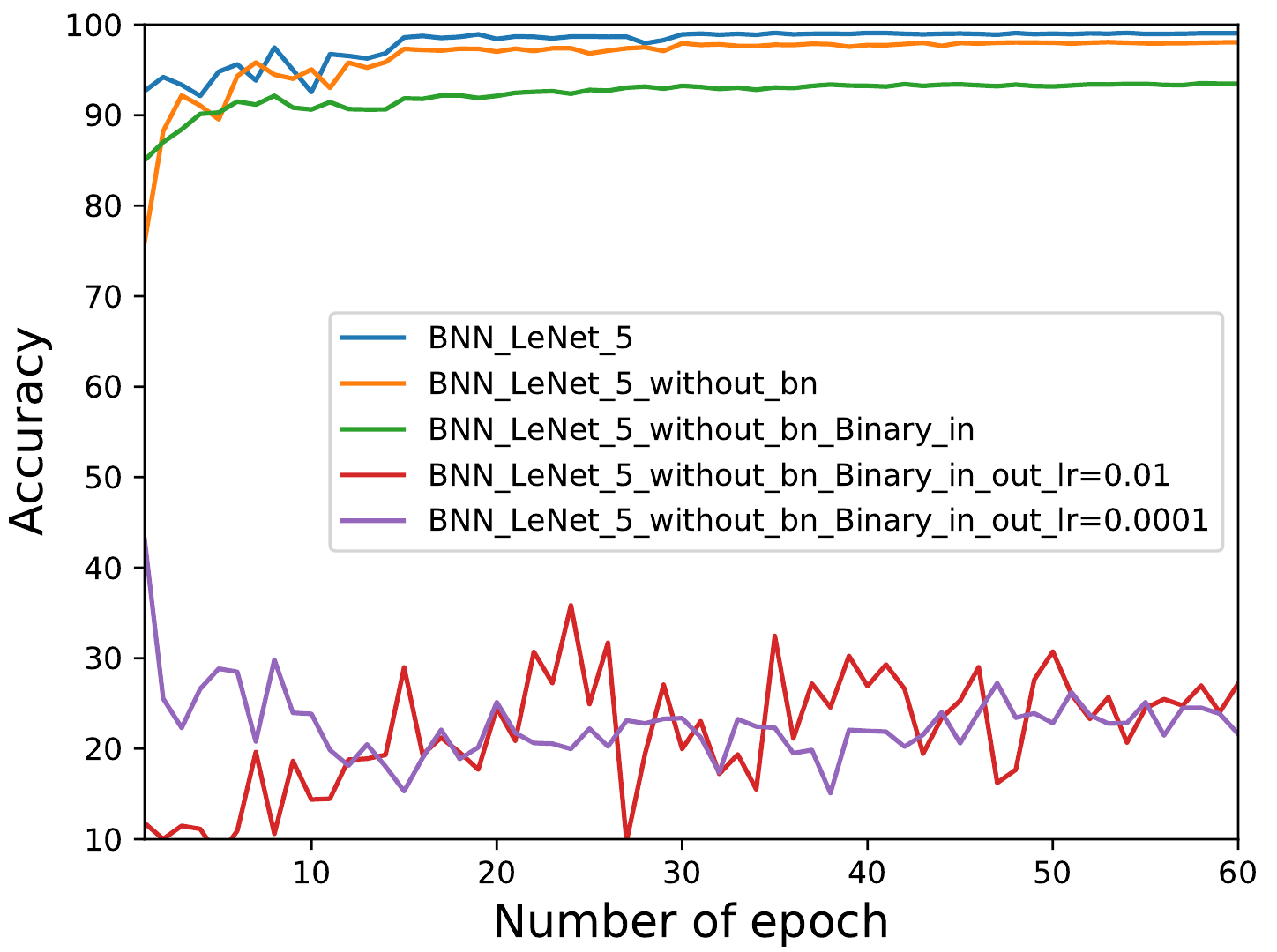}
  \caption{Accuracy of binarized LeNet5 network and the impact of input/output layer binarization as well batch normalization (bn) on accuracy loss  \vspace{-0.4cm}}
  \label{fig:LeNet_5}
\end{figure}

Although binarization enhances the system's efficiency in terms of memory usage, energy, and performance, it usually comes at the cost of accuracy loss compared to its high-precision counterpart. Therefore, using proper methods and algorithms to preserve the accuracy of the network as high as possible is essential. Each iteration of training a network can be divided into three steps; forward pass, backward propagation, and parameter update. The weights during backward propagation and forward pass are binarized while keeping high precision weights during parameter update is necessary. Since parameter changes obtained by gradient descent are tiny, binarization ignores these changes and the network cannot be trained \cite{xnor_net, binaryconnect}. In addition, binarizing the input and output layer usually results in a huge accuracy loss. Figure \ref{fig:LeNet_5} depicts the accuracy of the binarized LeNet5 network trained for the MNIST dataset. This clearly shows the impact of binarizing the input and output layers as well as batch normalization (bn) on the accuracy of the network.

\section{Memristor-based accelerators for BNN}
\label{sec:SoTA}

Memristor crossbar arrays are tailored to perform analog VMM with more significant energy efficiency compared to their digital counterpart (CPU/GPU) \cite{VMM}. Although some memristor devices can potentially be programmed to multi-resistance levels, they have higher reliability, stability, and accuracy when fewer resistance levels are used. Hence, BNN-based applications where the main kernel is binarized VMM are the promising targets to be implemented using memristor devices. A small-scale demonstration of BNN on memristor devices is presented in \cite{10by10BNN} with focusing mainly on device variation and its effect on BNNs. A new methodology is proposed in \cite{BNN_more_active_row} to make the design more tolerant against device variations to be able to activate more word-lines and perform more computation at the same time.
Based on Equation \ref{eq:sign}, BNNs require signed representation, but negative numbers cannot be directly stored in memristors. Considering that, the existing BNN accelerators can be classified into two categories based on the solution that they employ.  

\begin{figure}[t]
\centering
  \includegraphics[width=\linewidth]{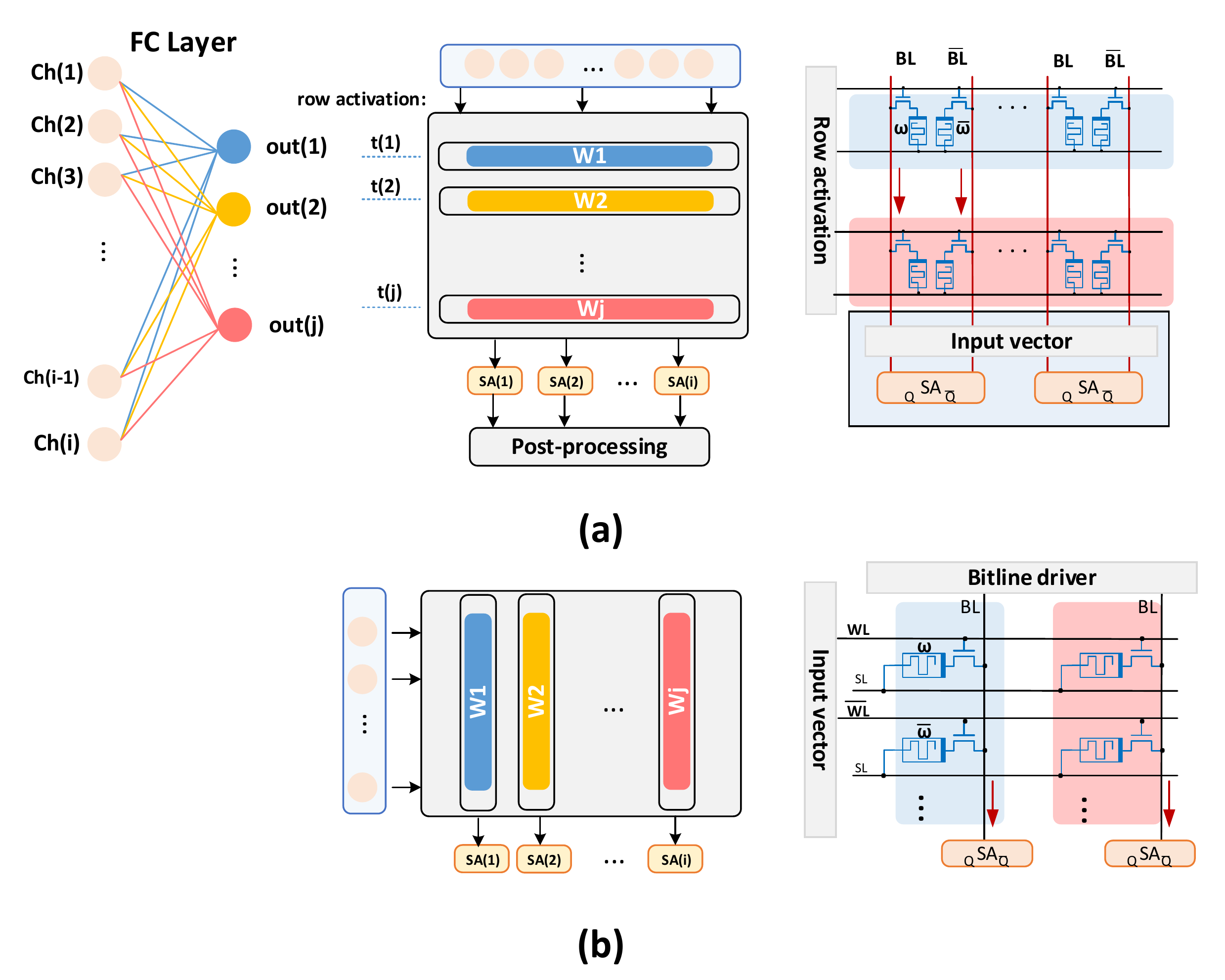}
  \caption{(a) BNN implementation using differential sensing and sequential XNOR operation \cite{BNN_differential_SA} (b) proposed design where massive XNOR operations are performed in parallel \vspace{-0.4cm}}
  \label{fig:FCLayer}
\end{figure}
\begin{itemize}[leftmargin=*]
\item Hardware solutions:
In order to deal with signed numbers, the weights and activation values can be converted each to two vectors containing only positive values \cite{highly_parallel_NN}. The two vectors of the weights and activation values are downloaded to the corresponding memristors and crossbar's input ports. Subsequently, the four possible partial results are computed. This requires a high number of memristor devices which translates to low area and energy efficiency. In addition, more complex input drivers are required to provide current in both directions. A similar approach is mapping positive and negative weights into different crossbars \cite{BCNN_RRAM,configurable_neurons}. Besides, ADC is exploited to compute the partial result when a BNN layer size is larger than the crossbar size. However, using ADCs imposes significant energy and area overhead to the system. An interesting approach is using one- \cite{BNN_ref_column1} or two-column reference memristors \cite{highly_robust} while the weights and activations are presented as \{0,1\}. In this design, the current flowing through the reference column(s) has to be mirrored equal to the number of columns in the crossbar. This increases the design complexity and energy consumption of the system. In addition, when a layer size cannot fit into a crossbar, it gets critical to have a flexible referencing scheme to avoid accuracy loss. We discuss this more in Section \ref{Section: accuracy}. \\

\item Algorithmic solutions:  
Binary multiply and accumulate operation can be replaced by \textbf{XNOR+popcount+post processing}. As a result, the weights and activations for BNN can be presented as unsigned \{0,1\} values. As a consequence and considering memristor crossbars, it makes the implementation simpler, but additional digital processing has to be done after the crossbar. 
Content-addressable memory (CAM) structure based on binary XNOR operation is used for BNN \cite{BNN_SA_is_memristor}. In this design, the activation function is implemented by a memristor where its state determines the input value for the
next layer. However, this suffers from an extremely high number of device programming which causes challenges in terms of reliability, performance, and energy. An XNOR-based robust design to device imperfections is proposed using a differential sensing mechanism \cite{BNN_differential_SA}. This design is closest to this work and is considered for our baseline. Figure \ref{fig:FCLayer}(a) illustrates how a fully connected layer is mapped to a crossbar. Due to the structure of the crossbar and mapping of the weights, outputs are generated sequentially. Besides, additional digital processing is required to generate the final result. \\ \vspace{-0.4cm}
\end{itemize}
\noindent
Considering the limitations and challenges of existing works, a high- performance and energy-efficient design of BNN is highly demanded. 


\section{Methodology}
\label{section:Methodology}
In this section, first, we discuss the implementation of BNNs on memristor crossbars based on XNOR operation. Second, we explain how the crossbar's input buffer containing activation values is managed to minimize data transfer between crossbars. \vspace*{-0.4cm}

\begin{figure*}[t]
\centering
  \includegraphics[width=\linewidth]{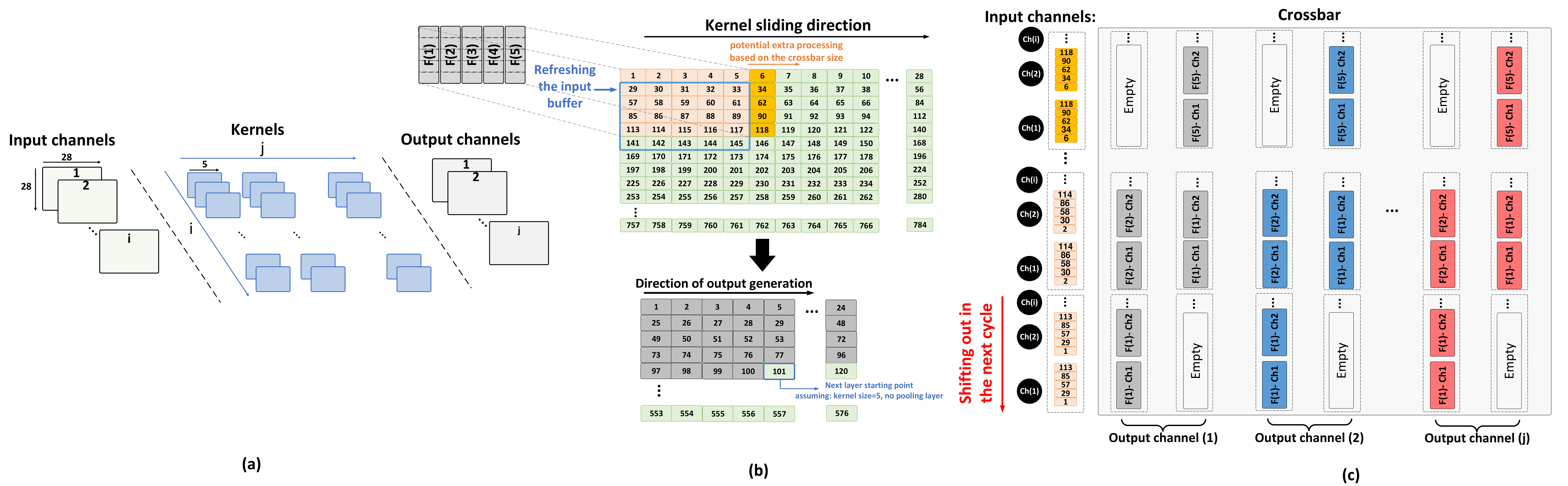}
  \caption{(a) Example of a CNN layer (b) details of a convolution operation with $5\times5$ and $28\times28$ kernel and input size (c) mapping of the activation values to the input buffer and kernels to the crossbar based on the proposed approach to minimize data transmission between layers by only streaming the newly computed activation values into the input buffer}
  \label{fig:CLayer}
\end{figure*}

\subsection{Proposed BNN implementation}
The multiply-accumulate operation between two signed binarized vectors can be replaced by \textit{XNOR} and popcount operations \cite{xnor_net}. To achieve that, first, the vectors are converted to unsigned where `-1' is represented as `0'. This is helpful since it simplifies the mapping of weights to the crossbar without concern for negative values. Second, by applying Equation \ref{eq:pop}, the final value is obtained where \textit{A'} is the unsigned representation of vector \textit{A}. In this equation, popcount() returns the number of ones in a bitstream and \textit{`vector size'} is the length of the two vectors. In the following, an example is provided to have better clarification. 

\begin{equation}
    \label{eq:pop}
    A*B = 2*Popcount(A^{'} \odot B^{'}) - vector\;size
\end{equation}

\noindent
$A=[1,-1,-1,1]\;B=[-1,1,1,1] \Rightarrow A*B=-2$\\
\noindent
$A^{'}=[1,0,0,1] \; B^{'}=[0,1,1,1] 	\Rightarrow A^{'} \odot B^{'}=[0,0,0,1]$
$A*B = 2*Popcount(A^{'} \odot B^{'}) - vector\;size = -2$

By applying the above method to a fully connected layer, the process of generating the activation value for the next layer can be expressed as (similarly to a convolution layer):
\begin{equation}
  \label{eq:activation}
  out_m= Sign (2*\sum_{k=1}^{I} (Ch_{k} \odot \omega_{k,m}) - vector\;size)  
\end{equation}
where $Ch_{k}$ represents the activation value for the current
layer, $\omega_{k,m}$ is the weight related to the $k^{th}$ input and $m^{th}$ output, and I is equal to the number of inputs of the current layer. Figure \ref{fig:FCLayer}(b) depicts how the above equation is implemented on a crossbar. Despite the approach illustrates in Figure \ref{fig:FCLayer}(a), the summation in the equation is performed in an analog way in the crossbar. In this mapping, each column corresponds to one output of the layer and they can perform the operations in parallel. However, in order to generate the final value, besides the sign function, other operations have to be performed. As to achieve that, it may require to place ADC to generate the actual value of this analog summation and perform other operations in the periphery of the crossbar accordingly. However, by changing the sign operation to a comparator where its reference is obtained from Equation \ref{eq:SA}, the output can be efficiently computed.       

\begin{equation}
  \label{eq:SA}
  SA\; reference = vector\;size /2  
\end{equation}
Based on this approach, first, we maximized the number of parallel activation values that can be computed for a BNN layer. Second, to avoid reducing this efficiency by using a high-resolution ADC, a simple analog comparator with a smart referencing value is deployed. This not only performs the sign operation, but also omits extra digital processing in the periphery.

\subsection{Efficient data movement}

Data movement between the BNN layers may influence the performance and energy of the system \cite{ISAAC}, but is often overlooked by the existing works. In this subsection, we focus on how the data should be transferred from one convolutional layer to the next to minimize the number of transactions and the size of a buffer placed between layers. This approach can be utilized for both binary and non-binary datatypes.  \par

Figure \ref{fig:CLayer}(a) depicts an example of convolution layer where the kernel matrix is convolved into the \textit{``i''} input channels to generate data for the \textit{``j''} output channels. In this example, the input size for each channel and the kernel size are $28\times 28$ and $5\times 5$, respectively. Figure \ref{fig:CLayer}(b) illustrates the details of the convolution operation where each kernel slides on a corresponding input channel to produce the partial result. The kernels are programmed to the crossbar while the data of input channels corresponding to the current operating window (highlighted by light orange) are buffered and sent to the wordlines of the crossbar. When the operating window slides, the data has to be sent and reorganized in the buffer to be matched to the weights of the kernel programmed into the crossbar. However, bringing the whole data again for the next operating window is not an efficient way since most of it already exists in the input buffer of the crossbar from the previous operating window. \par

To provide better data utilization and reduce the number of transactions, Figure \ref{fig:CLayer}(c) demonstrates an efficient mapping of kernels in the crossbar as well as activation value in the input buffer. In this approach, the kernels and the input data within the operating window are sliced into columns. The same columns for different input channels are packed together and placed in the input buffer. The next columns are stacked on top of each other as highlighted by the light orange color in Figure \ref{fig:CLayer}(c). The kernels are also treated the same way. By doing that, when the operating window slides to the right (assuming stride is one), the left-most columns for all the input channels are shifted out and new data corresponding to the right-most columns are streamed into the buffer. There is no need to change the mapping of the kernels in the crossbar and they always reside in front of the right inputs. When the operating window reaches the last columns, it has to be shifted down and starts from the most left column again. Therefore, the input buffer is refreshed and filled with data highlighted by the blue window in \ref{fig:CLayer}(b). As a result, maximum data is utilized when the operating window slides while the input buffer can be implemented as simple as possible.  \par

In order to maximize the performance, we can exploit parallelization and pipelining. In case the crossbar dimension is large enough, the computation for current and next operating windows can be performed in parallel. As illustrated in Figure \ref{fig:CLayer}(b) and (c), an extra column (highlighted by bright orange) required for the next operating window is placed into the input buffer of the crossbar. Besides, we have to consider another column in the crossbar to be able to generate the value for both operating windows simultaneously. It has to be taken into account that this extra input set should not contribute to the computation of the current window. Therefore, the memristors located in the first column and in front of this extra input set should be programmed to logic value `0'. It is worth mentioning that the kernels for other output channels are programmed to different columns of the crossbar to maximize parallelization. However, in case the crossbar has a lower number of columns, we need to deploy more crossbars to avoid an excessive number of reprogrammings. Besides parallelization, the same pipelining approach presented in \cite{ISAAC} can be applied in this work. Depending on the kernel size of the next layer in the network, when enough elements are produced for the output channels of the current layer, the operation can be started for the next layer.

\section{Intra-layer accuracy analysis}
\label{Section: accuracy}

\begin{figure}
\centering
  \includegraphics[width=0.7\linewidth]{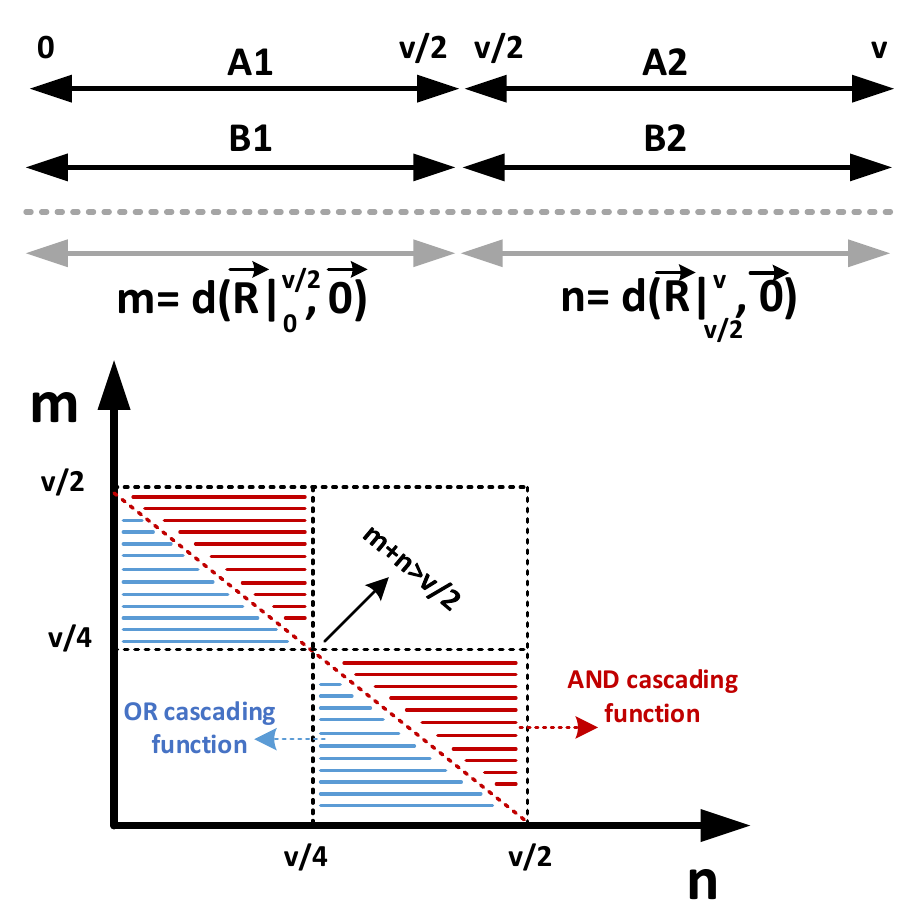}
  \vspace*{-0.2cm}
  \caption{Illustration of the regions where logical \textit{AND} and \textit{OR} functions inject inaccuracy into the network}
  \label{fig:math_illustrate}
\end{figure}

In Section \ref{section:Methodology}, the proposed implementation was presented where a single SA can generate the activation value for the next BNN layer (see Figure \ref{fig:FCLayer}). However, if the weights that are supposed to be in a single column of a crossbar cannot fit into it, they have to be split and mapped to more columns.  
Therefore, the final activation value has to be calculated from the intermediate activation values obtained from different sets of columns. This is where inaccuracy is injected into the network with a certain probability. 
\noindent
In the following, the ideal situation is formulated where the crossbar size is equal or greater than the vector size. $\overrightarrow{A}$ and $\overrightarrow{B}$ are the two binary vectors and $d(\overrightarrow{R}, \overrightarrow{0})$ is the hamming distance between vectors $\overrightarrow{R}$ and  $\overrightarrow{0}$. \vspace{0.2cm}    

\noindent
\textit{Vector size} = $\nu$, \textit{Crossbar size} = \textit{$C\ge \nu$} \\
input 1: $\overrightarrow{A}$, input 2: $\overrightarrow{B}$ \\
\noindent $\overrightarrow{R}= \overrightarrow{A} \bigodot \overrightarrow{B}$  \\
$out_{golden}( \overrightarrow{R}) = \left\{ \begin{array}{rcl}
1 & \mbox{if}
& d(\overrightarrow{R}, \overrightarrow{0}) > \nu/2  \\ 0 & \mbox{otherwise}
\end{array}\right.$ \\

In case the crossbar size is not big enough, the formulation is changed as presented below. As an example, we assume the crossbar size is half of the vector size. Therefore, each vector has to be split into two parts and mapped to two columns of the crossbar.  \vspace{0.2cm}     

\noindent \textit{Vector size} = $\nu$, \textit{Crossbar size}: \textit{$C= \nu/2$} \vspace{0.2cm}  \\
input 1: $\overrightarrow{A}|_{0}^{\nu/2}, \overrightarrow{A}|_{\nu/2}^{\nu}$ where $\overrightarrow{A} = [\overrightarrow{A}|_{0}^{\nu/2}, \overrightarrow{A}|_{\nu/2}^{\nu}]$\\ 
input 2: $\overrightarrow{B}|_{0}^{\nu/2}, \overrightarrow{B}|_{\nu/2}^{\nu}$ where $\overrightarrow{B} = [\overrightarrow{B}|_{0}^{\nu/2}, \overrightarrow{B}|_{\nu/2}^{\nu}]$ \vspace{0.2cm} \\

\noindent $\overrightarrow{R}|_{0}^{\nu/2}= \overrightarrow{A}|_{0}^{\nu/2} \bigodot \overrightarrow{B}|_{0}^{\nu/2}$ $\;\; \; \overrightarrow{R}|_{\nu/2}^{\nu}= \overrightarrow{A}|_{\nu/2}^{\nu} \bigodot \overrightarrow{B}|_{\nu/2}^{\nu}$  \vspace{0.2cm} \\

\noindent$out_{p1}( \overrightarrow{R}|_{0}^{\nu/2}) = \left\{ \begin{array}{rcl}
1 & \mbox{if}
& d(\overrightarrow{R}|_{0}^{\nu/2}, \overrightarrow{0}) > (\nu/2)/2  \\ 0 & \mbox{otherwise}
\end{array}\right.$ \\
\noindent$out_{p2}( \overrightarrow{R}|_{\nu/2}^{\nu}) = \left\{ \begin{array}{rcl}
1 & \mbox{if}
& d(\overrightarrow{R}|_{\nu/2}^{\nu}, \overrightarrow{0}) > (\nu/2)/2  \\ 0 & \mbox{otherwise}
\end{array}\right.$ \\

\noindent Since we mapped the vector into two columns, two intermediate activation values ($out_{p1},out_{p2}$) are obtained. The final value depends on the ``\textbf{cascading function}''. This function can be a simple logical \textit{AND} or \textit{OR} function.  \vspace{0.2cm} \\
\noindent $out(\overrightarrow{R}|_{\nu/2}^{\nu},\overrightarrow{R}|_{0}^{\nu/2})= out_{p2}( \overrightarrow{R}|_{\nu/2}^{\nu}) \wedge out_{p1}( \overrightarrow{R}|_{0}^{\nu/2}) $ \vspace{0.2cm} \\
In the case of logical \textit{AND} as an example, the following conditions show the scenarios where the output of cascading function differs from the golden output. This is also illustrated in Figure \ref{fig:math_illustrate}. The axes are the hamming distance obtained from the result of the first and second parts of the output vector. The red and blues regions indicate where the \textit{AND} and \textit{OR} functions generate inaccurate results. Following are the conditions where the output of \textit{AND} cascading function differs from the golden output. \vspace{0.2cm}  \\ 
\noindent $out(\overrightarrow{R}|_{\nu/2}^{\nu},\overrightarrow{R}|_{0}^{\nu/2}) \neq  out_{golden}( \overrightarrow{R}) $ if:  \\
$
\left\{ \begin{array}{rcl}  
d(\overrightarrow{R}|_{0}^{\nu/2}, \overrightarrow{0})+ d(\overrightarrow{R}|_{\nu/2}^{\nu}, \overrightarrow{0}) > \nu/2 \\d(\overrightarrow{R}|_{0}^{\nu/2}, \overrightarrow{0}) < \nu/4
\end{array}\right.$  \\ \indent \large \textbf{$\vee$} \\ 
$\left\{ \begin{array}{rcl}  
d(\overrightarrow{R}|_{0}^{\nu/2}, \overrightarrow{0})+ d(\overrightarrow{R}|_{\nu/2}^{\nu}, \overrightarrow{0}) > \nu/2 \\d(\overrightarrow{R}|_{\nu/2}^{\nu}, \overrightarrow{0}) < \nu/4
\end{array}\right.$ \\ 

\begin{figure}
\centering
  \includegraphics[width=0.7\linewidth]{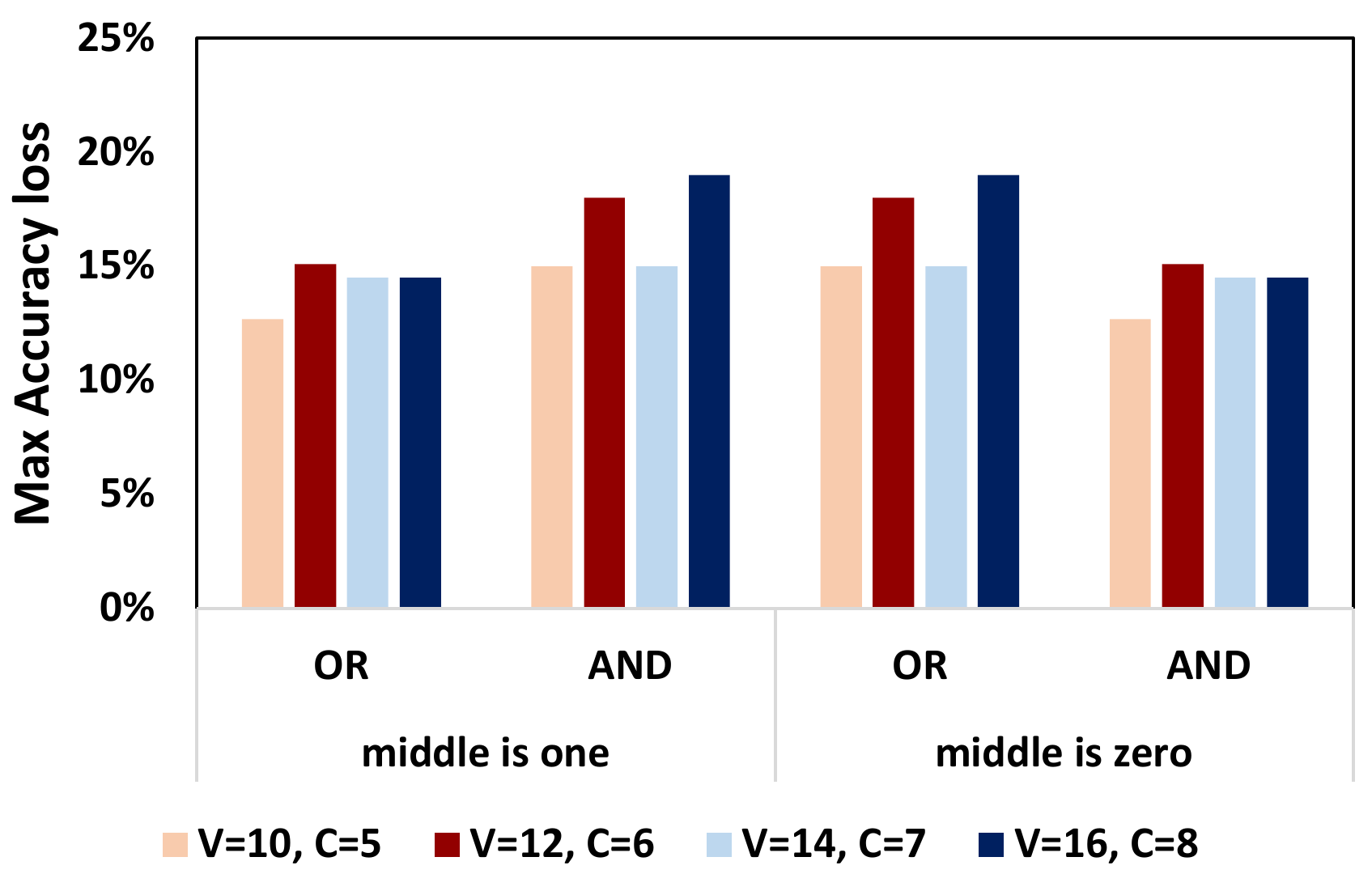}
  \caption{Maximum accuracy loss simulated for all possible input vectors for different vector sizes (V), crossbar size (C), and cascading functions}
  \label{fig:accuracy_loss}
\end{figure}

\noindent The number of input vectors that fall in these regions (blue or red in Figure \ref{fig:math_illustrate}) are calculated based on Equation \ref{eq:number_of_cases}. According to this equation, Figure \ref{fig:accuracy_loss} depicts the maximum accuracy loss for two cascading functions considering two boundary conditions. This is done by generating all the input sets to verify the Equation \ref{eq:number_of_cases}. We observe that the accuracy loss does not have considerable changes over vector sizes as the relative area associated to inaccurate region remains the same (Figure \ref{fig:math_illustrate}).   

\begin{equation}
\label{eq:number_of_cases}
  \begin{split}
& \forall (m, n) \in Solution \; Set: \\
& \#(\overrightarrow{A}, \overrightarrow{B}) = ({}_mC_{\nu/2}*2^m*2^{\nu/2-m})* \\ & ({}_nC_{\nu/2}*2^n*2^{\nu/2-n}) 
  \end{split}
\end{equation}

\begin{figure}
\centering
  \includegraphics[width=0.9\linewidth]{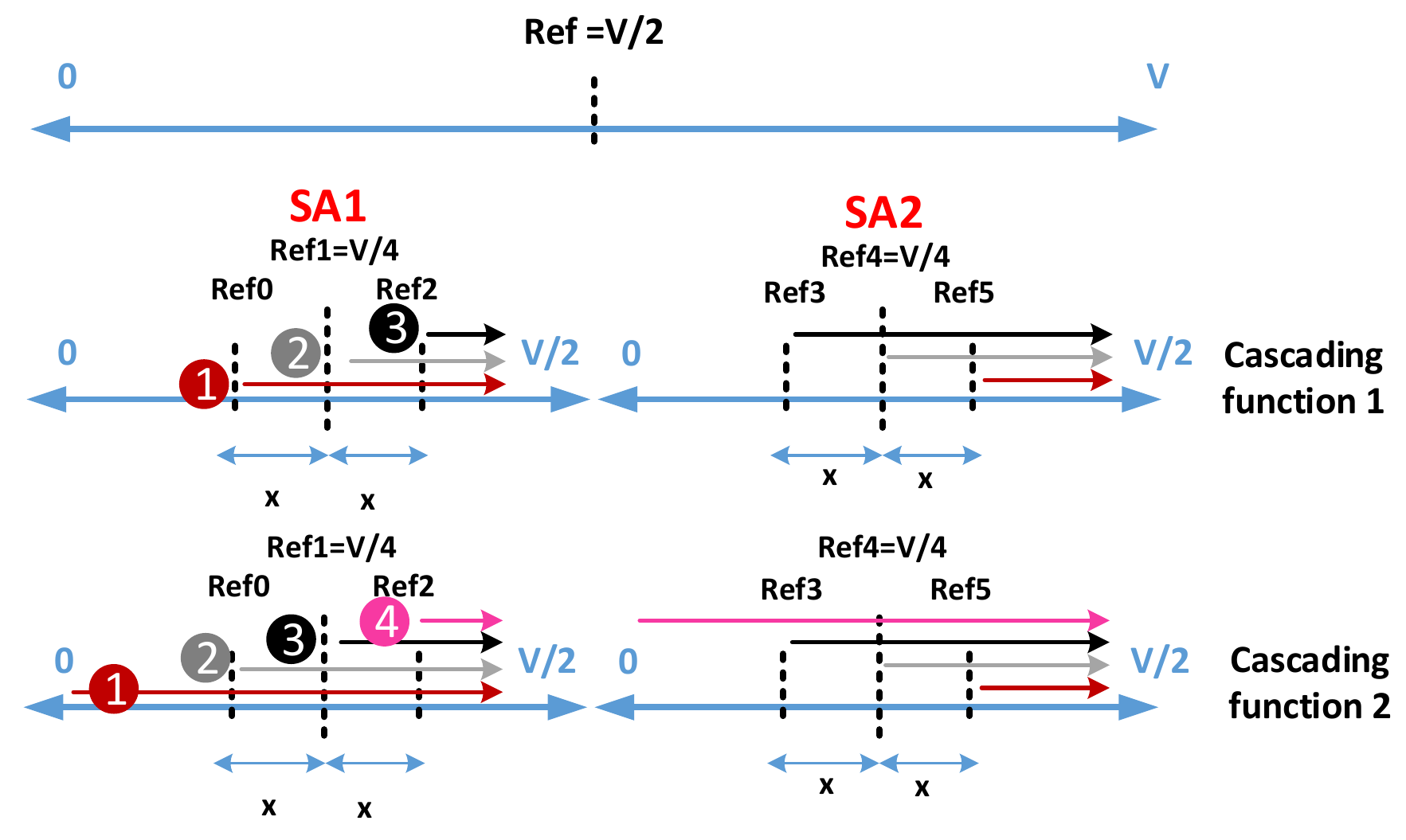}
  \caption{Illustration of two cascading functions where two auxiliary references added to the main reference}
  \label{fig:Cascading}
\end{figure}

To reduce the accuracy loss, more references can be considered. This leads to more intermediate results which provide us with more information as well as more flexibility to have more advanced cascading functions. However, we should take into account that keeps adding references increases the hardware complexity of SA. In the following, we investigate the implication of the number of references as well as their actual values on accuracy loss. Figure \ref{fig:Cascading} presents an example where two auxiliary references are added to the main reference. In this scenario, three intermediate values are produced for each of the output vectors and the final activation value should be decided based on them. We illustrate
two possible cascading functions in this figure. The first function comprises three conditions, where meeting each, can set the final activation value to one. These are based on the fact that the summation of two hamming distances obtained from two output vectors should be greater than half of the original vector size (Equation \ref{eq:SA}). This function always set the activation value to one accurately (true positive), while it misses to set it to one in some cases (false negative). Considering that, the second cascading function makes the conditions more relaxed. The probability of accuracy loss for these two functions is computed in the following.         

\begin{figure}
\centering
  \includegraphics[width=\linewidth]{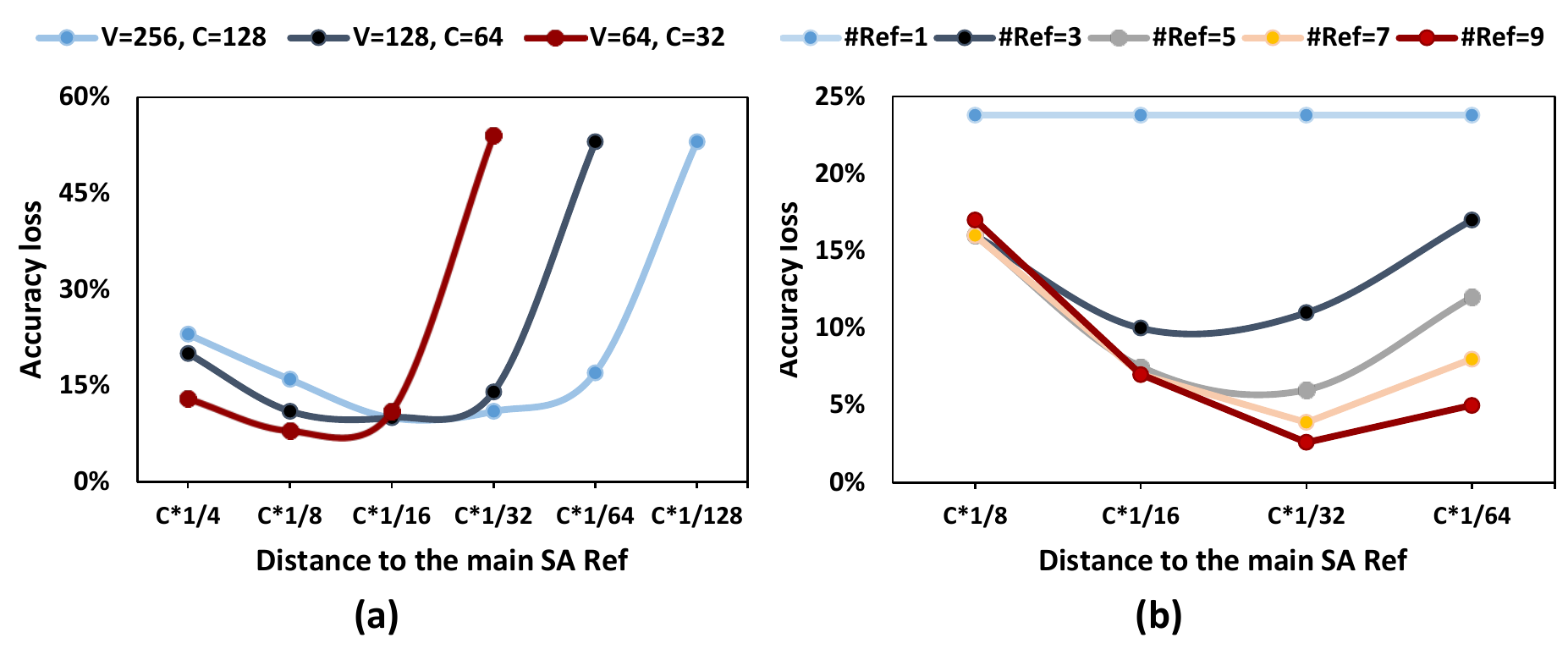}
  \caption{(a) Accuracy loss based on the distance of two auxiliary references to the main reference (b) effect of number of auxiliary references on accuracy}
  \label{fig:accuracy_loss_random_vector_size_Ref_combined}
\end{figure}


\begin{figure}[]
\centering
  \includegraphics[width=0.5\linewidth]{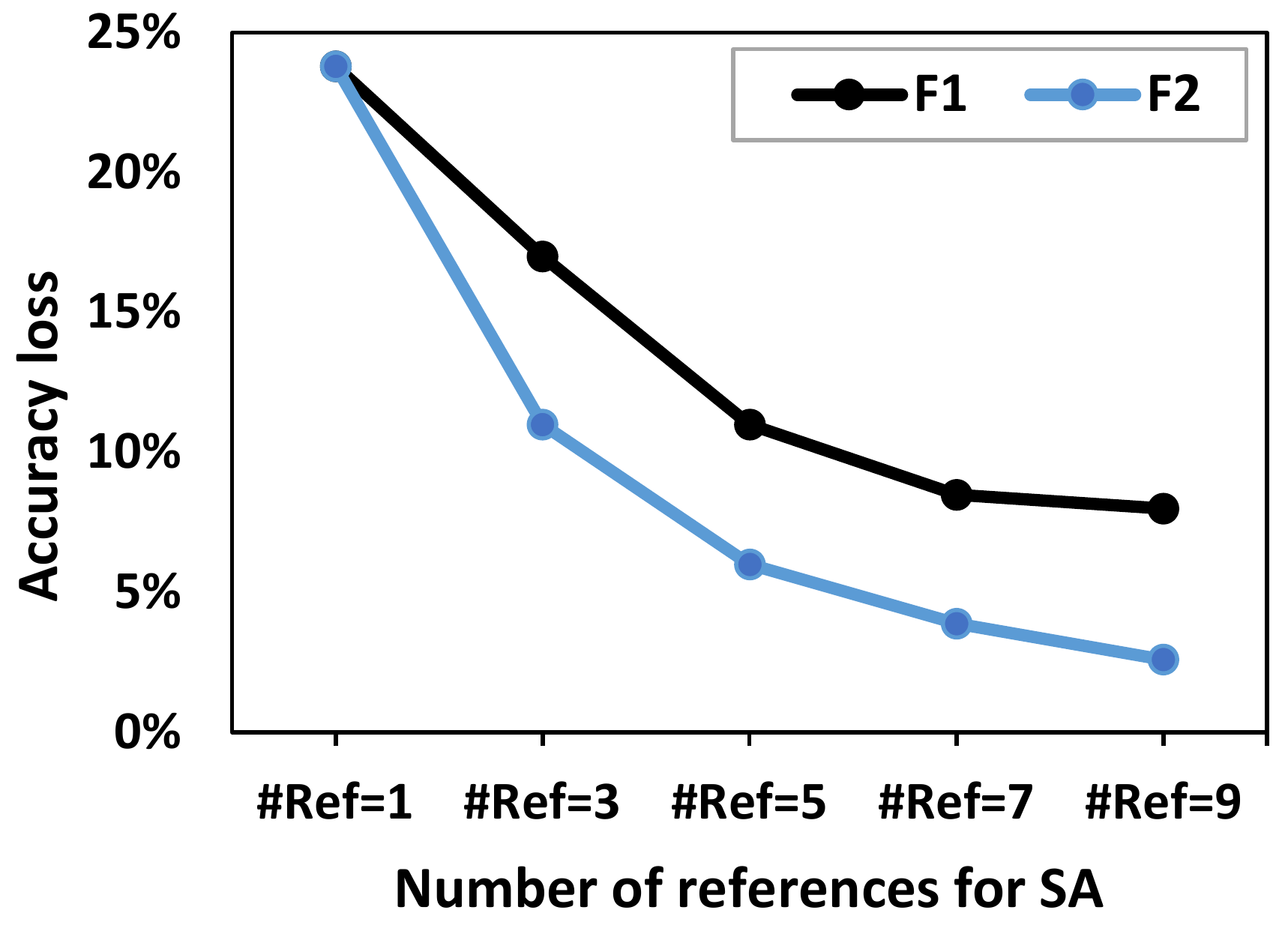}
  \caption{ Impact of the two cascading functions (illustrated in Figure \ref{fig:Cascading}) on accuracy loss  \vspace{-0.4cm}}
  \label{fig:Function_cascading}
\end{figure}

\begin{table*}[t]
\tiny
\centering
\caption{Typologies of the BNNs and their software accuracy}
\label{tab:benchmarks}
\resizebox{0.9\textwidth}{!}{%
\begin{tabular}{|l|l|l|c|}
\hline
Name & \multicolumn{1}{c|}{Topology} & Dataset & Accuracy \\ \hline
LeNet-5 & 5x5,6 - 2x2 Pool - 5x5,16 - 2x2 Pool - FC(120) - FC(84) - FC(10) & MNIST & \%98 \\ \hline
CNN-1 & 5x5,5 - 2x2 Pool - FC(720) - FC(70) - FC(10) & MNIST & \%97 \\ \hline
CNN-2 & 7x7,10 - 2x2 Pool - FC(1210) - FC(1210) - FC(10) & MNIST & \%98 \\ \hline
MLP-S & FC(784) - FC(500) - FC(250) - FC(10) & MNIST &  \%97 \\ \hline
MLP-M & FC(784) - FC(1000) - FC(500) - FC(250) - FC(10) & MNIST & \%98.2 \\ \hline
MLP-L & FC(784) - FC(1500) - FC(1000) - FC(500) - FC(10) & MNIST & \%98.4  \\ 

\hline
\end{tabular}%
}
\end{table*}

\noindent
$P_{Loss}(F1)= \\ \textbf{P}([SA_{out2}>Ref5 \wedge SA_{out1}<Ref0] \wedge [SA_{out2}+SA_{out1}>\nu/2]) + \textbf{P}([SA_{out2}<Ref3 \wedge SA_{out1}>Ref2] \wedge [SA_{out2}+SA_{out1}>\nu/2]) + \textbf{P}([Ref4<SA_{out2}<Ref5] \wedge [Ref0<SA_{out1}<Ref1] \wedge [SA_{out1}+SA_{out2}>\nu/2]) + \textbf{P}([Ref3<SA_{out2}<Ref4] \wedge [Ref1<SA_{out1}<Ref2] \wedge [SA_{out1}+SA_{out2}>\nu/2])$ \\
\noindent
$P_{Loss}(F2)= \\ \textbf{P}([SA_{out2}>Ref5] \wedge [SA_{out2}+SA_{out1}<\nu/2]) + \textbf{P}([SA_{out2}>Ref5] \wedge [SA_{out2}+SA_{out1}<\nu/2]) + \textbf{P}([Ref4>SA_{out2}>Ref3] \wedge [Ref2>SA_{out1}>Ref1] \wedge [SA_{out2}+SA_{out1}<\nu/2]) + \textbf{P}([Ref5>SA_{out2}>Ref4] \wedge [Ref1>SA_{out1}>Ref0] \wedge [SA_{out2}+SA_{out1}<\nu/2])$

An important parameter that has a remarkable impact on the accuracy loss is the distance of auxiliary references to the main reference (\textit{``x''} in Figure \ref{fig:Cascading}). This is quite dependent on the distribution of data. Hence, the designer can analyze the network and based on that find the proper value for the references where the accuracy loss is minimized. Figure \ref{fig:accuracy_loss_random_vector_size_Ref_combined}(a) demonstrates the impact of this parameter for cascading function 2 assuming normal distribution. This is presented for different crossbar sizes (``\textit{c}''). The distance to the main reference is shown relative to the crossbar size. The figure indicates the importance of the values for the references and how considerably they can change the accuracy loss.  Another important parameter is the number of references. The implication on accuracy can be comprehended from Figure \ref{fig:accuracy_loss_random_vector_size_Ref_combined}(b). It is observed that by keep adding more references, an improvement in accuracy is reduced while more complexity is added to the hardware. Finally, the impact of the cascading functions on accuracy is evaluated in Figure \ref{fig:Function_cascading} over a different number of references. The same two methods presented in Figure \ref{fig:Cascading} are also used for the situation where we have more than three references. The figure indicates that choosing a proper function can help the accuracy of the system remarkably.  \vspace{-0.4cm}

 \section{Evaluation}
\label{section:Evaluation}

\subsection{Simulation setup}

Our simulation results are obtained by creating our PyTorch-based platform \cite{Simulator}. This platform is able to evaluate the accuracy, energy, and latency of different networks containing binarized and non-binarized layers. The software is written in a modular way to flexibly change network structure as well as different circuit-level parameters. The system runs at a clock frequency of 1GHz. The data bus between the crossbars has 32-bit width. Based on the 32nm technology node, transferring data to store it in an input buffer consumes 5mW \cite{PUMA,Mahdi_arch}. The energy and latency number of the ``Shift and Add'' unit required for non-binarized layers taken from \cite{Mahdi_arch}. In all the simulations, the crossbar size is $512 \times 512$ \cite{IBM_512_crossbar}. We use an analytical model based on a small PCM prototype and extend the memory to the required size. The model is acquired from the results of the EU project MNEMOSENE \cite{mnemosene}. Finally, the specification of ADC is taken from \cite{HDC}. \par

Our benchmark (MlBench) comprises 6 BNNs for machine learning applications. The structure of each network is listed in Table \ref{tab:benchmarks}. LeNet-5, CNN-1, and CNN-2 are convolutional networks, and MLP-S/M/L are multilayer perceptrons (MLPs) with different network scales \cite{PRIME}. These networks are evaluated on the widely used MNIST database of handwritten digits. We compare our design with a recent work published in one of the leading journals in this field \cite{BNN_differential_SA}. For this work, we instantiate the digital post-processing units (popcount) for every 16 columns of the crossbar instead of sequentially operating over all the columns (see Figure \ref{fig:FCLayer}(a)). This diminishes the latency overhead of digital processing for the baseline.  \vspace{-0.4cm}


    
    

\subsection{Results}
\textbf{Accuracy analysis}\\ 

\begin{figure}[t]
\centering
  \includegraphics[width=0.8\linewidth]{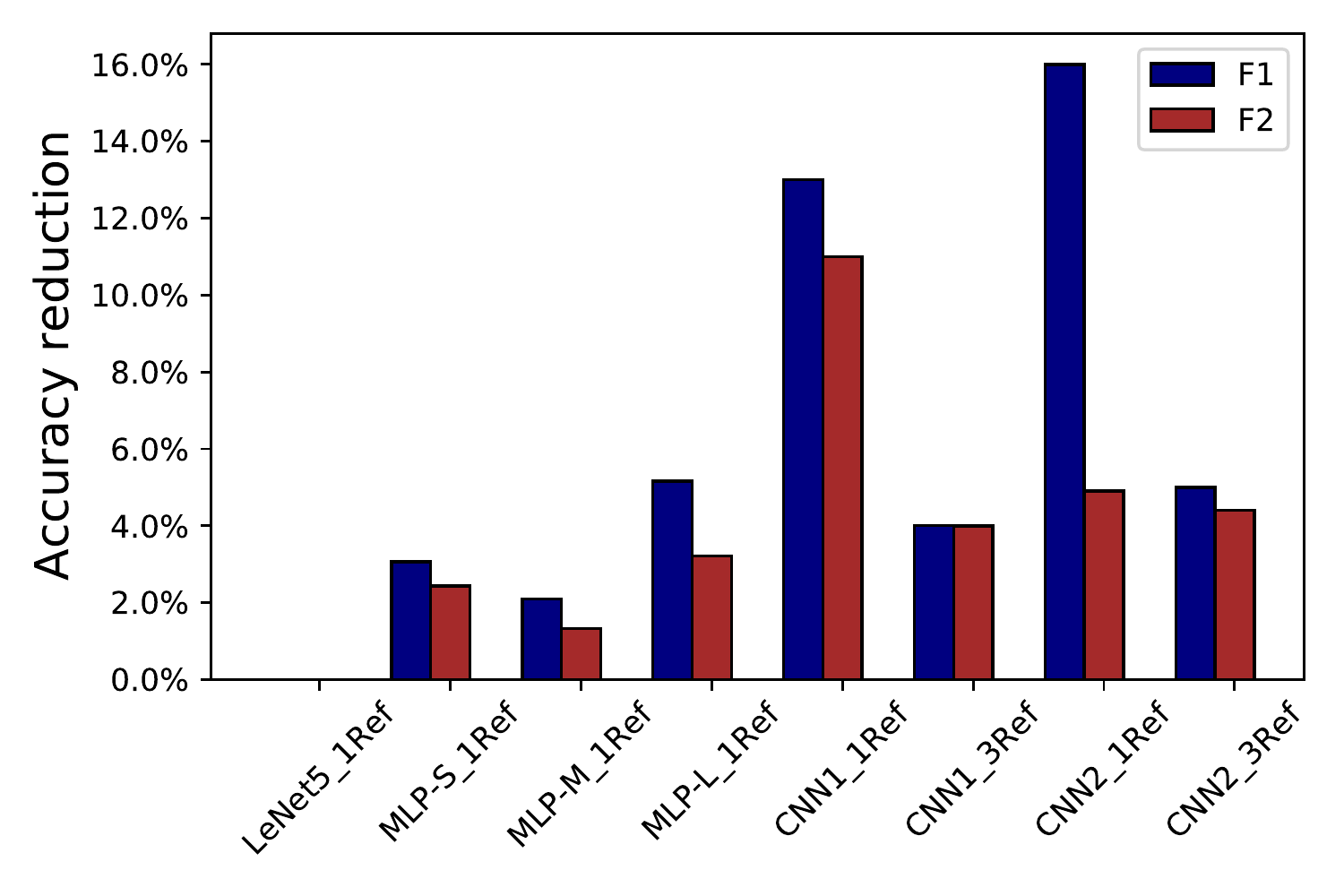}
  \vspace*{-0.4cm}
  \caption{Accuracy reduction for different network structures due to the crossbar size limitation and breaking the vectors over more crossbars}
  \label{fig:final_acc}
\end{figure}

Figure \ref{fig:final_acc} depicts the accuracy loss using our proposed approach compared to the software implementation. The figure presents the results for all the benchmarks considering two different cascading functions (see Figure \ref{fig:Cascading}). Since the size of each layer in LeNet-5 network is smaller or in the range of crossbar size, no accuracy loss is observed. However, this is not the case for the rest of the networks. The results show the importance of cascading function. As calculated in \ref{Section: accuracy}, ``F2'' is superior than ``F1'' due to less noise injection per layer. However, the difference depends on the network structure and distribution of data.  

\begin{figure}[t]
\centering
  \includegraphics[width=0.6\linewidth]{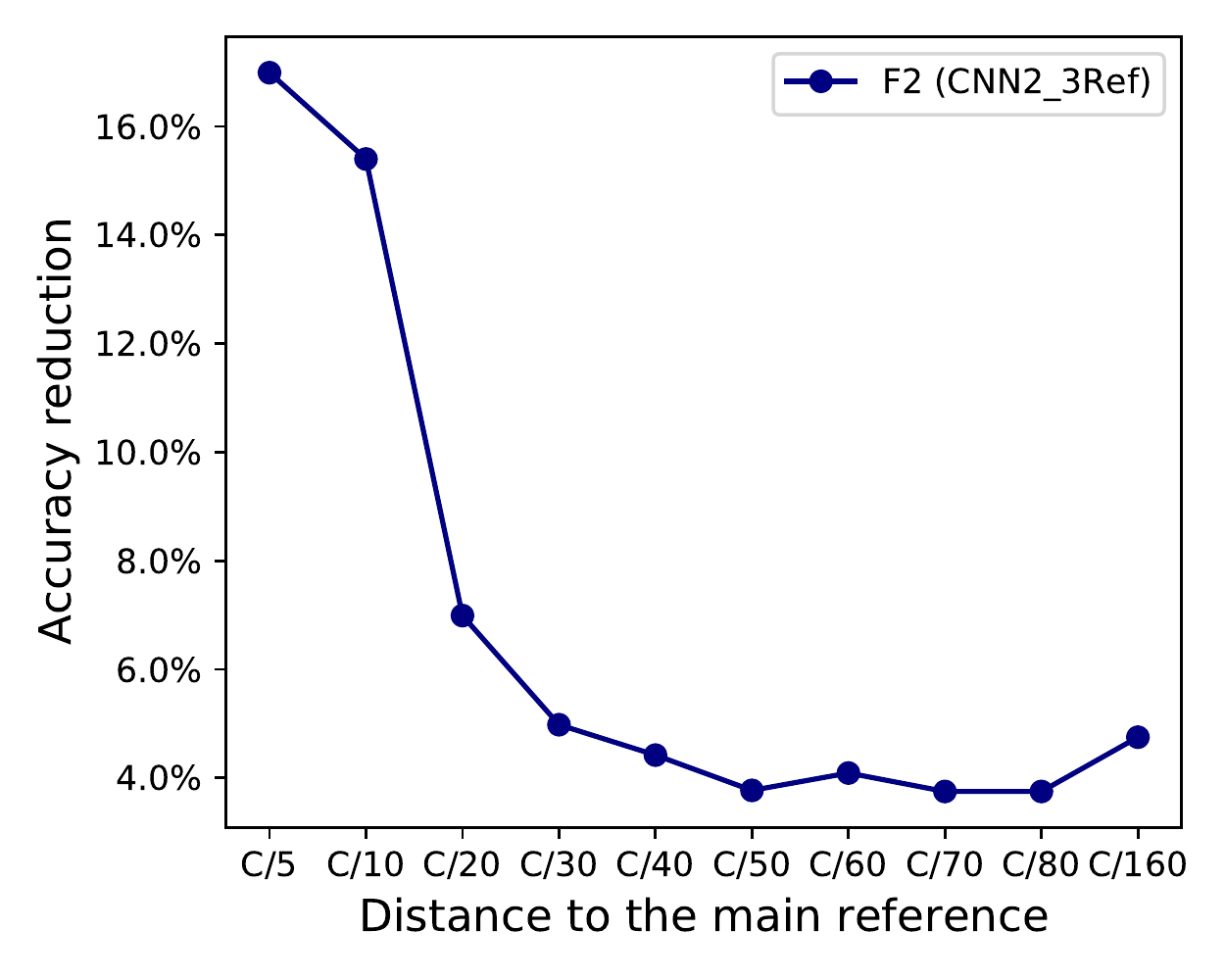}
  \vspace*{-0.2cm}
  \caption{Impact of auxiliary references and their distance from the main reference on accuracy loss \vspace{-0.4cm}}
  \label{fig:distance_acc}
\end{figure}

Other important parameters which can have a remarkable impact on accuracy are the number of references and their distance from each other. We ran the simulation for \textit{CNN1} and \textit{CNN2} networks with 3 references. As expected and can be seen in Figure \ref{fig:final_acc}, the presence of two auxiliary references helps to generate less inaccuracy in the network. Besides, Figure \ref{fig:distance_acc} depicts the consequence of distance between main reference and auxiliary references (``x'' in Figure \ref{fig:Cascading}). The distance is relative to the crossbar size (``C''). Placing the references far from or close to each other reduces their efficiency in eliminating the cases where inaccurate activation values are generated. Therefore, the designer should find the optimal value for the references.

\begin{figure}[t]
\centering
  \includegraphics[width=0.9\linewidth]{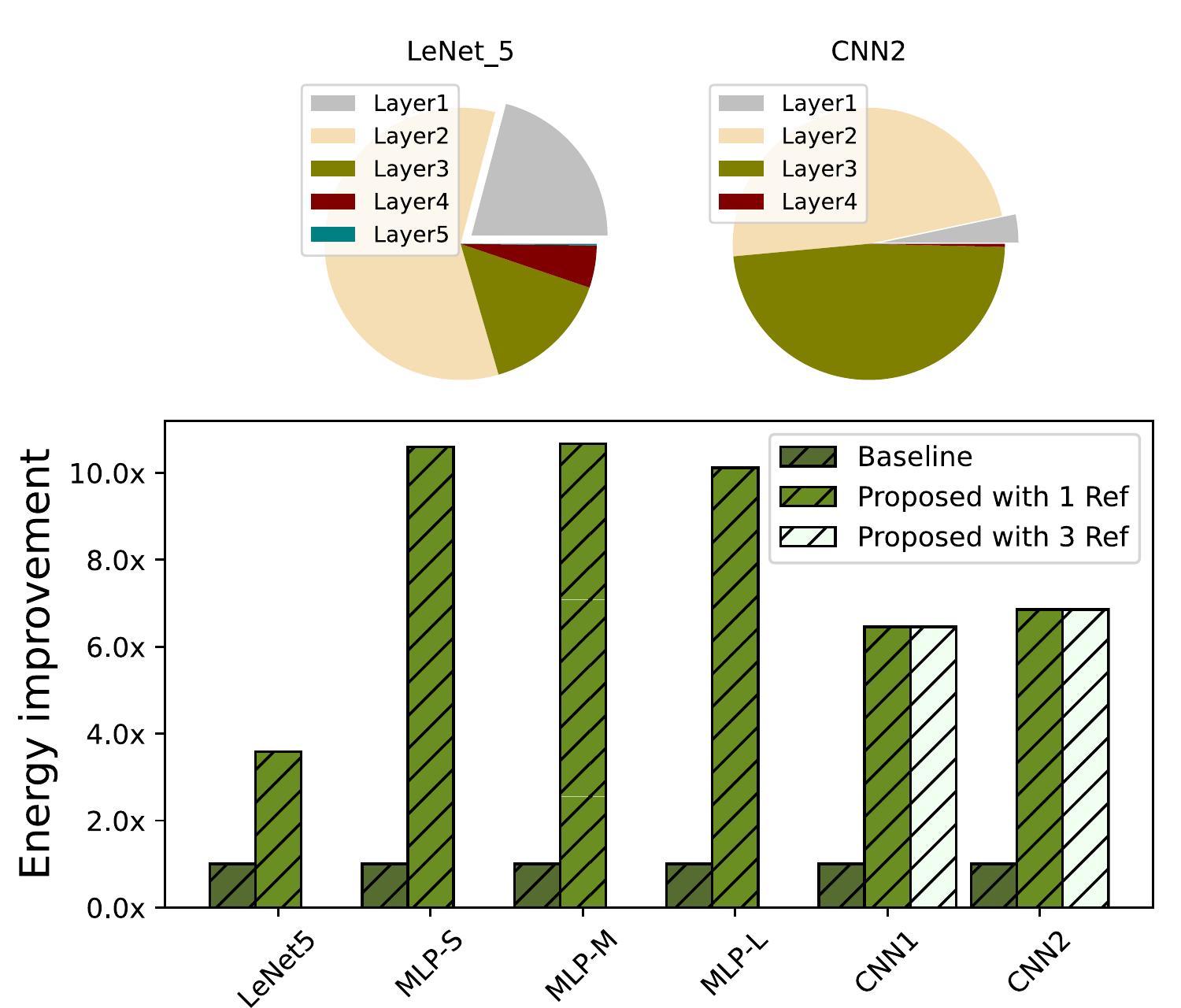}
  \vspace*{-0.2cm}
  \caption{Energy improvement compared to the baseline and break down of energy for different layers of two networks \vspace{-0.4cm}}
  \label{fig:energy_improv}
\end{figure}

\begin{figure}[t]
\centering
  \includegraphics[width=0.9\linewidth]{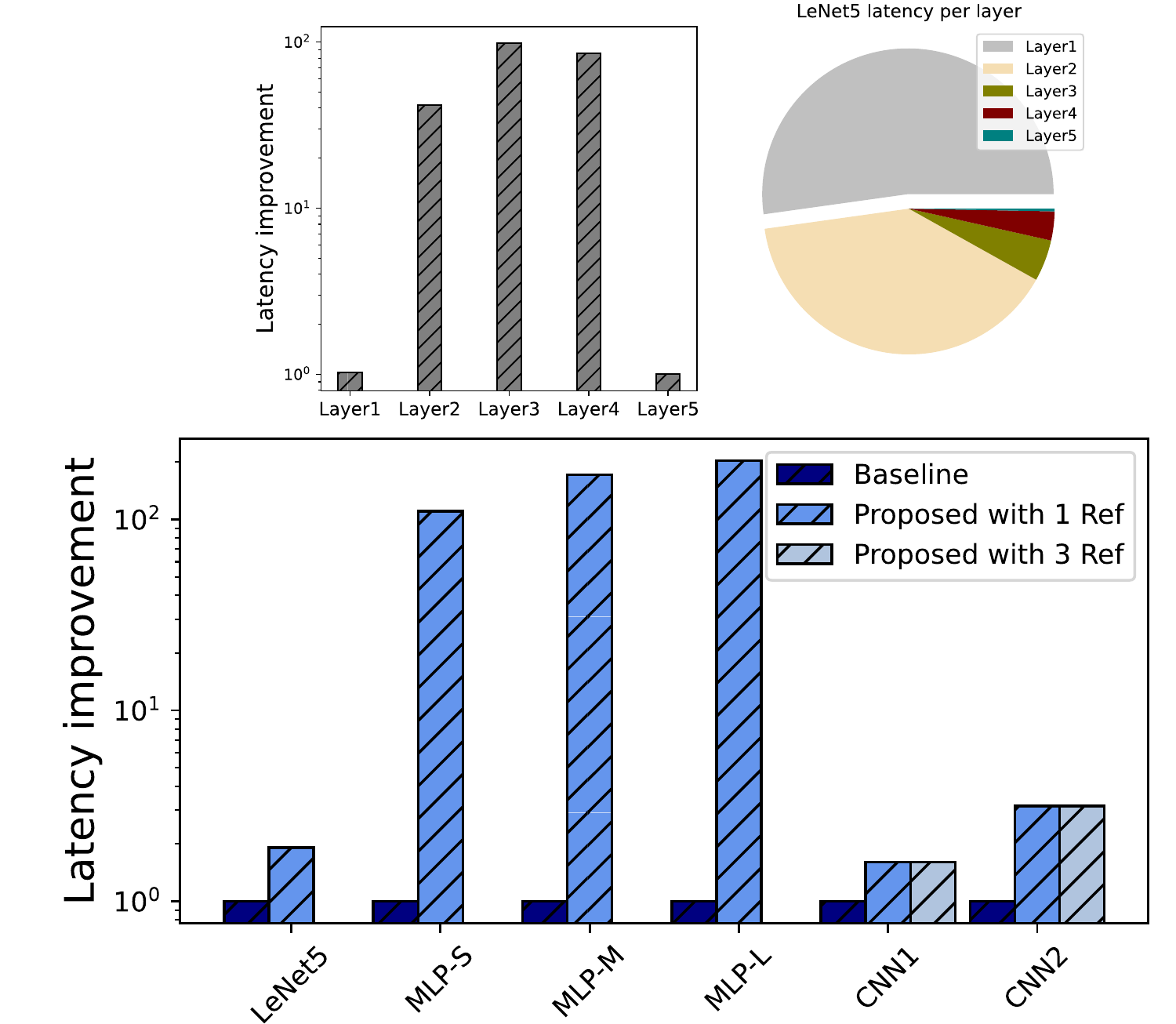}
  \vspace*{-0.2cm}
  \caption{Latency improvement compared to the baseline and its break down for different layers of LeNet-5 network  \vspace{-0.5cm}}
  \label{fig:latency_improv}
\end{figure}

\textbf{Energy and latency analysis}\\ 
Figure \ref{fig:energy_improv} depicts the energy improvement and the contribution of layers in total energy consumption considering two networks. The result shows that up to 10$\times$ improvement is achieved compared to the baseline. Energy improvement mainly is eventuated from less crossbar activation. In convolutional networks, since the first layer, which is not binarized, has the most contribution to the total energy, less improvement is obtained. In addition, a SA with three references requires three cycles to generate the output which leads to approximately three times more energy consumption. However, the impact on the total energy of the network is negligible due to the small contribution of SAs.            

Figure \ref{fig:latency_improv} shows the latency improvement of the entire network. The remarkable improvement is obtained mainly due to computing each activation value in a non-sequential manner as well as computing the activation values among different output channels in parallel. Similar to the energy number, the improvement for convectional networks is less due to the large contribution of the first layer to the total latency of the network. Therefore, as a solution to reduce the overhead of this layer on the network, a designer may allocate more resources for this layer to compute the activation values for different operating windows in parallel. However, in our simulation, we consider as minimum as possible resources for each layer.     
\section{Conclusion and future direction}
\label{section:Conclusion}

This paper proposed a novel in-memory memristor-based design that substantially improves the performance and energy efficiency of BNN applications. The proposed XNOR-based BNN design, replace the functionality of ADC and post-processing with a SA while maximizing parallelization and resource utilization in the design with a novel mapping of weights and activation values in the crossbar and its input buffer. The design can outperform the baseline specifically in intermediate layers. On average this work is able to yield 8$\times$ and 60$\times$ higher energy and performance than the baseline. In our future work, we consider the impact of variability in references on the accuracy of the network. Besides, we evaluate the design for larger and more complex networks to comprehend the impact of inaccuracy injected into intermediate layers on the final accuracy of the networks.

\bibliographystyle{IEEEtran}
\bibliography{ref}

\end{document}